\documentclass[preprint,12pt]{elsarticle}
\usepackage{graphics}
\usepackage{rotating}
\usepackage{mathrsfs}
\usepackage{amsmath}
\usepackage{lscape}
\usepackage[usenames,dvipsnames]{color}
\usepackage{pgfplots}
\usepackage[utf8]{inputenc}
\usepackage{graphicx} 
\usepackage{amsmath,amssymb}
\usepackage{natbib}
\usepackage[utf8]{inputenc}
\usepackage{graphicx} 
\usepackage{amsmath,amssymb}
\usepackage{multirow}

\DeclareUnicodeCharacter{00A0}{ }
\begin{document}

\title{Odd-even staggering in neutron drip line nuclei}
\author{S. A. Changizi}%
\author{Chong Qi}%
\ead{chongq@kth.se}
\address{Department of Physics, Royal Institute of Technology (KTH), SE-10691 Stockholm, Sweden}

\begin{abstract}
We have done systematic Hartree-Fock-Bogoliubov calculations in coordinate space on the one-quasi-particle energies and binding energy odd-even staggering (OES) in semi-magic nuclei with the zero-range volume, mixed and surface pairing forces in order to explore the influence of their density dependence. The odd-$N$ isotopes are calculated within the blocking scheme. The strengths for the pairing forces are determined in two schemes by fitting locally to reproduce pairing gap in $^{120}$Sn and globally to all available data on the OES of semi-magic nuclei with $Z\geq8$. In the former calculations, there is a noticeable difference between the neutron mean gaps in neutron-rich O, Ca, Ni and Sn isotopes calculated with the surface pairing and those with the mixed and volume pairing.
 The difference gets much smaller if the globally optimized pairing strengths are employed. The heavier Pb isotopes show the opposite trend. Moreover, large differences between the mean gap and the OES may be expected in both calculations when one goes towards the neutron drip
 line.
\end{abstract}

\maketitle

\section{Introduction}

The study of highly unstable nuclei around the neutron drip
line has been one of the most active research fields in nuclear physics in the past few decades (see Refs. \cite{Tani13,Mich09,nature486} and references therein). The properties of those isotopes heavier than oxygen are mostly unknown and rely on the theoretical extrapolation based
on models optimized for known nuclei. The predictions are also important inputs for the study of the astrophysical r-process.
A challenging problem that one realizes is the strong deviation between theory and experiment and between different models that may occur when one goes toward those extreme cases with large $N/Z$ ratio.
The deviation is mostly related to our limited understanding of the theoretical uncertainty and the underlying effective nuclear force \cite{0954-3899-41-7-074001,PhysRevC.89.054314,PhysRevC.71.054311, arXive.1501.03572,Dudek13}. Significant efforts have been devoted to optimize the single-particle spectroscopy and binding energy prediction of various mean-field models \cite{PhysRevC.89.054314,Xu2013247,Wu15,Goriely201568}. For examples, in recent studies, the uncertainties in predicting the location of neutron drip line as induced by the single-particle energy, pairing and the macroscopic symmetry energy at sub-saturation density are highlighted \cite{Afa15,Wang15,Chen15}. Extensive work has also been done in studying the pairing features of drip
line nuclei in relation to the coupling involving loosely bound orbitals as well as the continuum.
However, it is still difficult to constrain the property of the effective nuclear pairing interaction for which no universal form has been achieved.

The so-called density-dependent zero-range (or contact) pairing interaction has been widely used in nuclear mean field and density functional calculations due to its intrinsic simplicity. 
In its simplest form, the zero-range $\delta$ interaction is given
by (see, e.g., Ref. \cite{Dobaczewski1206} and references therein)
\begin{equation}
\label{eq:surface}
V_{pair}(\textbf{r},\textbf{r}^{\prime})= V_{0}\left(1-\eta\frac{\rho(\textbf{r})}{\rho_{0}}\right) \delta (\textbf{r}-\textbf{r}^{\prime})
\end{equation}
where $V_{0}$ is the pairing strength, $\rho(\textbf{r})$ is the isoscalar 
nucleonic density and $\rho_{0}=0.16 fm^{-3}$. $\eta=0$, 0.5 and 1 correspond to the so-called
volume, mixed and surface pairing interactions, respectively. As it is defined, the surface pairing generates a pairing field peaked around the
nuclear surface whereas the volume pairing is mainly active inside the nucleus.
One question is how the density dependence of the zero range pairing interaction affects the pairing correlation. 
To address this issue, extensive calculations have been done within the Hartree-Fock (HF) plus BCS
and Hartree-Fock-Bogoliubov (HFB) approaches.
A systematic calculation of the odd-even staggering (OES) in binding energies has been done in Ref. \cite{PhysRevC.79.034306} by using the HF+BCS approach with three different density dependent pairing forces as well as by using the HFB approach with the mixed pairing. No significant difference between different calculations was seen.
A mixed pairing force is used in the systematic studies of Refs. \cite{Wang15,nature486} and  Monte Carlo calculations in the configuration space (seniority-zero) in Refs. \cite{PhysRevC.83.014319,Vol15}. A surface pairing was used in recent calculations on the electric dipole strength of the nucleus $^{120}$Sn \cite{Roc15} and the soft dipole excitations in neutron-rich O, Ca and isotopes \cite{Mat05}. Calculations with the volume pairing were also done in Ref. \cite{Mat05}. No noticeable difference between calculations with the two types of pairing forces of surface and volume was seen.
HFB calculations for the isotopic chain $^{100-132}$Sn was given in Ref. \cite{PhysRevC.71.054303} where it was found that the  pairing density was insensitive to the density dependence of the
pairing force.
As commented in Ref. \cite{Dobaczewski1206}, one may not be able to extract reliable information on the density
dependence of the effective pairing interaction from
available data on the OES in nuclear binding energy.

One may expect that the different density dependence of the pairing force
may lead to drastic differences of pairing fields at the nuclear surface in very neutron rich nuclei where weakly bound orbitals are coupled to the continuum, e.g., when one goes towards neutron-rich Sn isotopes beyond $N=82$ \cite{Dobaczewski2001361,Dobaczewski20021521,Dobaczewski01032002,Bennaceur2002}.
In Ref.  \cite{PhysRevC.80.044328} the pairing
vibrations in $^{124,136}$Sn was analyzed with the HFB+QRPA approach where it was shown that neutron transition density and pairing vibration in the neutron-rich nucleus $^{136}$Sn is more sensitive to the density dependence of the pairing than that in $^{124}$Sn.  A slight preference for the surface-peaked pairing was suggested in  \cite{PhysRevC.79.034306} based on binding energy systematics. Calculations on the $\alpha$ clustering on the nuclear surface of heavy nuclei also favors surface-enhanced pairing interaction \cite{PhysRevC.81.064319,Ward2013}.
 The density dependence of the pairing may also influence the pair transfer properties of neutron Sn and light semi-magic neutron-rich nuclei \cite{PhysRevC.71.064306, PhysRevC.84.044317,PhysRevC.82.024318}. The effect of the density dependence of the pairing force was also studied on the odd-even staggering in the charge radii  in Ref. \cite{Fay00}, on the excitation energies and transition probabilities of the first $2^+$ states in Ref. \cite{Tolo11} and on the pair transfer probabilities of neutron-rich Cr isotopes in Ref. \cite{Grasso2013}. The transfer probabilities in those nuclei depend strongly on the density dependence and strengths of the pairing force and the persistence of the pairing correlation at drip line.
In Ref. \cite{PhysRevC.88.034314} HFB calculations were done with surface-peaked zero-range and finite-range pairing forces, where it suggests that pairing can even persist in nuclei beyond the drip line. Such pairing persistence can indeed be sensitive to the density dependence of the zero-range pairing \cite{PhysRevC.91.024305}. Systematic calculations over the nuclei chart in Ref. \cite{PhysRevC.91.024305} are done in the harmonic oscillator space. It is expected that a more precise integration of the HFB equation (see, e.g., Refs. \cite{Dob04,Sch15,Pei08}) and a reliable treatment of the continuum may be obtained by calculation in the coordinate space, which allows a proper description of the asymptotic behavior of the loosely bound quasi-particle orbital of drip line nuclei.
In those nuclei the spatial structure of the quasi-particle orbitals may be sensitive to both the single-particle energy and the pairing correlation.
 A related subject is the possible isospin dependence of the pairing force \cite{Mar07,Mar08,Yam09,Sag13,PhysRevC.91.047303,Yam12,Ber09,Ber12} and the OES of the binding energy \cite{PhysRevC.88.064329}, which has not been pinned down yet.

In this paper we are interested in examining systematically the effects of the density dependence of the pairing interaction on the OES in the binding energies of neutron-rich nuclei from calculations within the HFB approach in the coordinate space with a proper treatment of the continuum. 
The OES will be compared with the calculated mean gaps of the corresponding even-even nuclei. The so-called volume, surface and mixed pairing force will be used. We are also interested in seeing how the calculations are influenced by the different choice of pairing strength. For that purpose we have used two schemes to determine the strengths of the pairing forces by reproducing the neutron pairing gap in the nucleus $^{120}$Sn as well as by fitting to available data on the OES of semi-magic nuclei with $Z\geq8$.

The paper is organized as follows. In Sec. \ref{sec:MF}, we briefly discuss the HFB framework and the blocking scheme for odd-$A$ nucleus. Calculated results for both even-$N$ and odd-$N$ semi-magic Sn, Ni, Ca and Pb isotopes are analyzed in  Sec. \ref{sec:Res}. In Sec. \ref{var} we optimize the strength of the different pairing forces by fitting to the available experimental data on OES. 
Then we compare in detail calculations for the OES of neutron-rich nuclei with different density-dependent pairing forces and pairing strengths.
Finally a summary is given in Sec. \ref{sec:con}.

\section{Formalism}
\label{sec:MF}
\subsection{Theory}
The HFB approach with effective zero-range pairing forces is a reliable and computational convenient way to study the nuclear pairing correlations in both stable and unstable nuclei. 
The HFB framework has been extensively discussed in the literature \cite{ring2004nuclear,Dobaczewski1984103,PhysRevC.53.2809,bender2003self,Dobaczewski1206}  and will only be briefly introduced here for simplicity. 
In the standard HFB formalism, the Hamiltonian is reduced to
the mean field in the particle-hole channel and the pairing field in 
the particle-particle channel as
\begin{equation}
\label{eq:H}
 \begin{pmatrix}
 (H-\lambda) & \Delta \\
 -\Delta^{*} & -(H-\lambda)^{*}
 \end{pmatrix}\begin{pmatrix}
	 U_{k}(r)\\
	 V_{k}(r) 	\end{pmatrix}= E_{k} 
	\begin{pmatrix}
	 U_{k}(r)\\
	 V_{k}(r)
	\end{pmatrix} , 
\end{equation}
where $U_{k}$ and $V_{k}$ are the two components of the single quasi-particle 
radial wave functions. 
We have used $k = {\tau nlj}$ as a short-hand notation for the
quantum numbers of the system since we have assumed spherical symmetry where $n$, $l$, $j$ are the number of nodes, orbital angular momentum and total angular momentum, respectively. $\lambda$ is the chemical
potential. The self-consistent HFB equations allow one to compute energies and wave functions of the quasi-particles. The energy spectrum of HFB quasi-particles contains both discrete bound quasi-particle states with quasi-particle energy $E\leq \lambda$ and quasi-particle continuum with $E>\lambda$.

The local densities can be written
using the radial functions as
\begin{eqnarray}
\rho(r)&=&
  \frac{1}{4\pi r^2}\sum_{k}(2j_k+1)V_k^2(r) + [U_{\delta}^2(r)-V_{\delta}^2(r)]\nonumber\\
\tilde{\rho}(r)&=&
  \frac{1}{4\pi r^2}[\sum_{k}-(2j_k+1)U_k(r)
  V_k(r) +U_{\delta}(r)V_{\delta}(r)]~~~~
\end{eqnarray}
where $\delta$ denotes the state that is to be blocked. 
For a given odd-$A$ nucleus,
possible blocked configurations can be defined from the calculated low-lying
quasi-particle spectra in neighboring even-even nuclei and the occupation
probabilities of those orbitals. It may be useful to point out that, unlike the cases in deformed nuclei where all Nilsson orbitals are doubly degenerate if the effect of the time-odd field is omitted,  one can have more than one pair within a given single-particle orbital. When one pair of $m$-orbital $\delta=nljm$ and its time reversal is being blocked, the degeneracy of the blocked orbital effectively becomes $2j_{\delta}-1$.

There are intensive ongoing studies on the blocking calculations of odd-$A$ nuclei, especially in the intermediate-mass and heavy nuclei regions. For examples, deformed one-quasi-particle states  actinide and rare-earth nuclei were recently studied  within covariant density functional theory in Ref. \cite{Afanasjev2011177}.
The systematics of one-quasiparticle configurations in neutron-rich Sr, Zr, and Mo isotopes was studied within the axial Gogny HFB approach based on the D1S and D1M  forces by applying the equal filling approximation method in Ref \citep{PhysRevC.82.044318}. In Ref. \cite{Dobaczewski1504} the properties of nuclei in the nobelium region with $92 \leq Z \leq 104$ and $144\leq N \leq 156$ were calculated using the covariant, Skyrme and Gogny functionals. A self-consistent treatment of the blocking for the one-quasiparticle HFB state is done in Ref. \cite{PhysRevLett.113.162501} by taking into account beyond mean field effects using the  Generator Coordinate Method. Systematic calculations on the one-quasiproton excitations in the rare-earth region are done within the Skyrme HFB approach in Ref. \cite{PhysRevC.81.024316}. In an earlier paper \cite{Sakakihara2001649},  the charge radii as well as the odd-even staggering in isotope shifts of Sn, Ba, Yb, and Pb isotopes were calculated systematically within the Skyrme HFB approach. Extensive works have been done recently on blocking calculations  within the  BCS and HFB frameworks for binding energy calculations of odd-$A$ and odd-odd nuclei (see, e.g., Refs. \cite{PhysRevC.91.047303,PhysRevC.89.054314,PhysRevC.89.054320}). Most of above calculations are done within the harmonic oscillator instead of the coordinate space. The effects of pairing correlation on the neutron quasi-particle resonance in $^{47}$Si was analyzed recently in Ref. \cite{Kob15} in the coordinate space with scattering boundary condition. The  Woods-Saxon potential was used for the 
particle-hole channel.

\subsection{Numerical Implementation}
\begin{figure}  
\begin{center}
\includegraphics[width=0.4\textwidth]{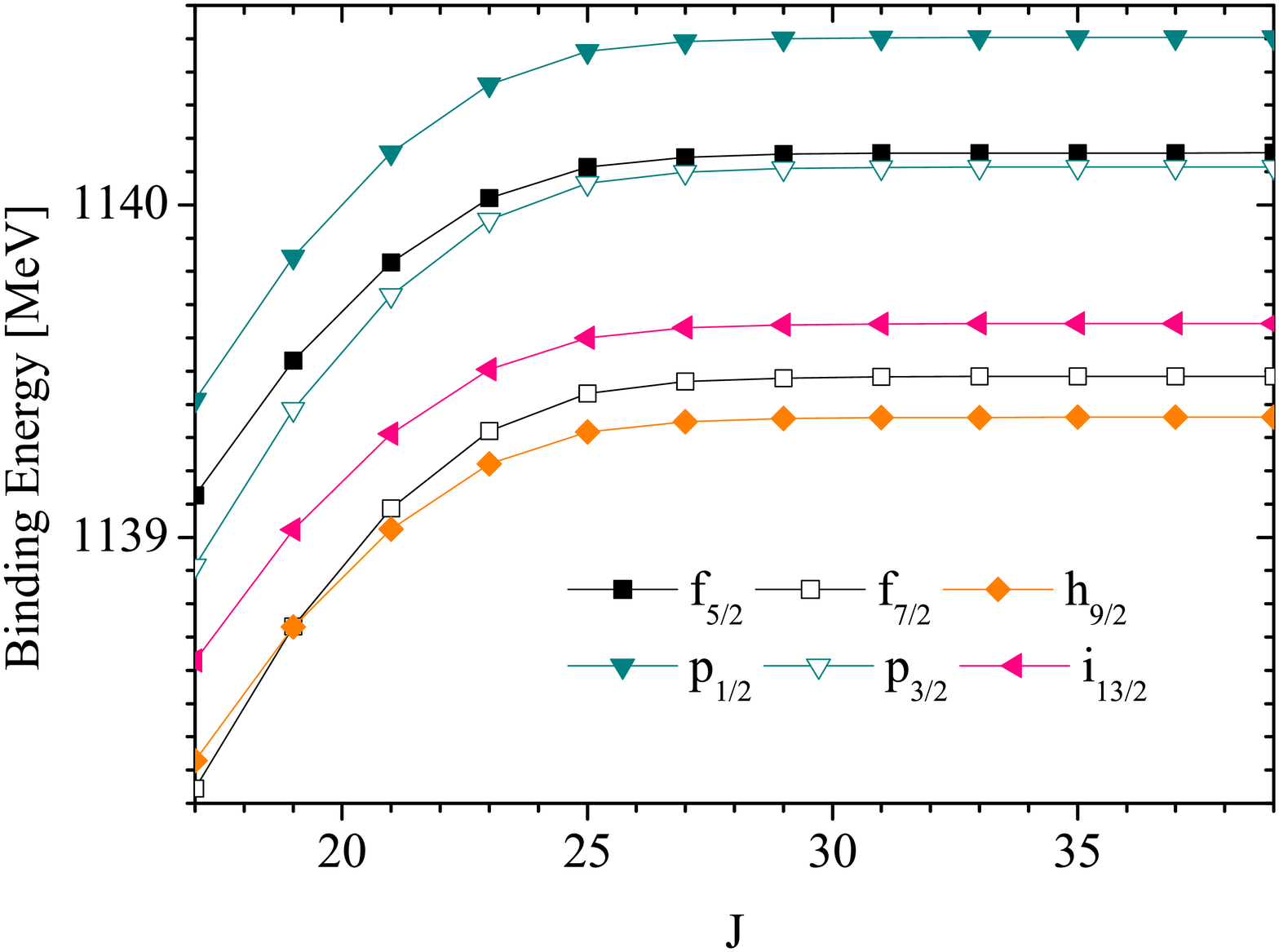}
\end{center}
\caption{\label{fig:Tin161_j} (color online) The convergence of the calculated ${}^{161}$Sn binding energy with different blocked levels as a function of $j_{max}$. Calculations are using the mixed pairing.
}
\end{figure}

We evaluated the 
coordinate-space solutions by using the HFB solver HFBRAD in a spherical box 
\cite{Bennaceur200596}. In that code the model space is confined by the maximum value for the angular momentum of the quasi-particle orbitals, which is taken as $j_{max}=25/2$ in the present work for all nuclei heavier than oxygen. We take $j_{max}=21/2$ instead for the very light O isotopes to avoid numerical instability. All calculations are practically converged with $j_{max}=25/2$ for nuclei that we are interested in. A typical example is given in Fig. \ref{fig:Tin161_j} where the binding energies of different one-quasiparticle states in the drip line nucleus $^{161}$Sn is calculated.  
We used the Skyrme 
functional with the standard SLy4 parameter set \cite{Chabanat1998} in the particle-hole channel  in that and all our following calculations. 
We have also performed calculations for all Pb isotopes with $j_{max}=39/2$ using all three pairing forces. Noticeable difference between calculations with $j_{max}=25/2$ and $j_{max}=39/2$  are only seen in the binding energies calculated using the surface pairing wheres the OES and mean gaps are not noticeably influenced.

It should be mentioned that we started this work by developing a blocking scheme for studying the properties of neutron-rich odd-$A$ nuclei. We realized at an early stage that it has been mostly implemented in the code HFBRAD, even though it is not mentioned in the original publication and, to our knowledge, the blocking scheme has not been tested in any earlier publication. In order to test the blocking algorithm within the HFBRAD algorithm, we compared its result with the prediction by the HFBTHO program \cite{Stoitsov20131592} since there is no other established blocking algorithm in coordinate space that is available to us. These two calculations do not necessary predict the same numerical results due to their different treatment of the nuclear continuum.
In both calculations the blocking is executed using the equal
filling approximation \cite{PhysRevC.78.014304,PhysRevC.81.024316}. The HFBTHO calculation is done in the configuration space defined by the harmonic oscillator with the frequency $\hbar\omega=1.2\times41/A^{1/3}$. We take in total $N_{max}=25$ major shells. 
The calculation is done by constraining spherical symmetry (to avoid the bias from spontaneous symmetry breaking). Our calculations with HFBTHO show that none of the Sn isotopes considered in the present work is significantly deformed. But in certain cases like $^{121}$Sn, the deformation effect can influence the binding energy by a few tens of keV.
Comparison between calculations with the earlier versions of the two codes for $^{120}$Sn and two other heavier even-even nuclei were done in Ref. \cite{Dob04} with the volume and mixed pairing interactions. The results are in general close to each other. But it should be noted that, in that calculation, the pairing strengths used in the two codes are different in order to get the same pairing gap for $^{120}$Sn (see also Ref. \cite{Pei08}). In Table \ref{tab:Sn120er} we have done similar comparisons for $^{120}$Sn with the three different pairing interactions. In all calculations the pairing strengths are determined by reproducing the pairing gap in $^{120}$Sn. For the HFBTHO calculations we have used both the normal harmonic oscillator (HO) basis and the transformed harmonic oscillator (THO) basis. The calculations are similar to each other in most cases. 
 In agreement with Ref. \cite{Pei08}, the absolute values of the pairing strengths fitted for the THO calculations are slightly smaller than those for HO and HFBRAD calculations for all three pairing forces employed.
In general the differences between the two HFBTHO and HFBRAD codes get smaller if the THO basis with improved description of the continuum was employed in the former case (see, discussions in Ref. \cite{Dob04} and references therein).
However, they still did not converge to the same results. In particular,  the THO calculations predict more neutron pairing energies than HFBRAD for all three interactions.
The largest deviations between the two codes are seen in calculations with the surface interaction where the neutron pairing energies differ by as much as 300-400 keV. This is related to the fact that the pairing wave functions are more dispersed for calculations with the surface interaction than those with the other two.

In Table \ref{tab:Sn121er} we compared 
three calculations for the nucleus $^{121}$Sn with the lowest quasi-particle orbitals $d_{3/2}$, $h_{11/2}$ and $g_{7/2}$ being blocked. Only calculations with the mixed pairing is shown for simplicity. We take the same pairing strength for both calculations.
In this case it is seen that the largest relative differences between HFBRAD and HFBTHO appear in the calculated quasi-particle energies and the neutron pairing energies. 
The neutron pairing gaps and the absolute values of the neutron pairing energies given by the THO calculations are slightly larger than those from HFBRAD in all three cases (with less than 2\% difference). In the former case, as in Table \ref{tab:Sn120er}, one needs a slightly smaller pairing strength in order to reproduce the same pairing gap as that of HFBRAD. But even in that case, the neutron pairing energy is still slightly overestimated.

\begin{sidewaystable}
\centering
\caption{\label{tab:Sn120er} Comparison between HFBRAD (with mesh size 0.1 fm) and HFBTHO calculations for the quasi-particle energies $E_{qp}$, neutron chemical potentials $\lambda_{n}$, neutron pairing energies $E^{n}_{pair}$, total radii, neutron kinetic energies $E^{n}_{kin}$, total spin-orbit energies $E^{tot}_{SO}$, direct Coulomb energies $E_{dir}$, and
total energies $E_{tot}$ in the nucleus $^{120}$Sn with different pairing interactions fitted to the average neutron pairing gaps $\Delta_{n} $ of $^{120}$Sn. HFBTHO calculations are done with both the HO and the THO bases.}
\resizebox{\linewidth}{!}{%
\begin{tabular}{l|rrrrrrrrrrrrrrr}
\hline
\hline
\multirow{2}{*}{} & \multicolumn{5}{c|}{Volume pairing interaction} & \multicolumn{5}{c|}{Mixed pairing interaction} & \multicolumn{5}{c}{Surface pairing interaction} \\ \cline{2-16}
                 & \multicolumn{1}{c|}{HO}     & \multicolumn{1}{c|}{THO}   & \multicolumn{1}{c|}{HFBRAD} & \multicolumn{1}{c|}{\multirow{2}{*}{diff HO}} & \multicolumn{1}{c|}{\multirow{2}{*}{diff THO}} & \multicolumn{1}{c|}{HO}    & \multicolumn{1}{c|}{THO}     & \multicolumn{1}{c|}{HFBRAD} & \multicolumn{1}{c|}{\multirow{2}{*}{diff HO}} & \multicolumn{1}{c|}{\multirow{2}{*}{diff THO}} & \multicolumn{1}{c|}{HO}    & \multicolumn{1}{c|}{THO}   & \multicolumn{1}{c|}{HFBRAD} & \multicolumn{1}{c|}{\multirow{2}{*}{diff HO}} & \multicolumn{1}{c}{\multirow{2}{*}{diff THO}} \\ \cline{1-4} \cline{7-9} \cline{12-14}
Parameters        & \multicolumn{1}{c|}{189.95} & \multicolumn{1}{c|}{188.8} & \multicolumn{1}{c|}{189.19} & \multicolumn{1}{c|}{}                         & \multicolumn{1}{c|}{}                          & \multicolumn{1}{c|}{289.1} & \multicolumn{1}{c|}{286.949} & \multicolumn{1}{c|}{287.85} & \multicolumn{1}{c|}{}                         & \multicolumn{1}{c|}{}                          & \multicolumn{1}{c|}{521.4} & \multicolumn{1}{c|}{513.4} & \multicolumn{1}{c|}{517.5}  & \multicolumn{1}{c|}{}                         & \multicolumn{1}{c}{}                          \\ \hline
$E_{qp}$          & 1.203                       & 1.200                      & 1.202                       & -0.001                                        & 0.002      & 1.298                      & 1.297                        & 1.298                       & 0.000                                         & 0.001                                & 1.771                      & 1.779                      & 1.774                       & 0.003                                         & -0.005                                 \\
$\lambda_{n}$     & -7.992                      & -7.993                     & -7.993                      & -0.002                                        & 0.000                                          & -8.014                     & -8.014                       & -8.015                      & -0.001                                 & -0.001                                         & -7.993                     & -7.994                     & -7.994                      & -0.001                              & 0.000                                          \\
$E^{n}_{pair}$    & -11.158                     & -11.242                    & -11.200                     & -0.043                                        & 0.042                                          & -13.587                    & -13.728                      & -13.659                     & -0.073                             & 0.069                                          & -27.827                    & -28.600                    & -28.161                     & -0.333                           & 0.439                                          \\
$\Delta_{n} $     & 1.310                      & 1.310                     & 1.310                      & 0.000                                         & 0.000 & 1.310                     & 1.310                       & 1.310                      & 0.000                                         & 0.000                               & 1.310                     & 1.310                     & 1.309                      & -0.001                                         & -0.001                                 \\
Radius            & 4.671                       & 4.671                      & 4.671                       & 0.000                                         & 0.000        & 4.674                      & 4.674                        & 4.674                       & 0.000                                         & 0.000                                & 4.687                      & 4.687                      & 4.687                       & 0.000                                         & 0.000                                  \\
$E^{n}_{kin}$     & 1340.994                    & 1341.117                   & 1341.177                    & 0.183                                         & 0.060                                          & 1342.017                   & 1342.160                     & 1342.203                    & 0.186                        & 0.044                                          & 1345.629                   & 1346.009                   & 1345.865                    & 0.236                      & -0.145                                         \\
$E^{tot}_{SO}$    & -48.552                     & -48.583                    & -48.566                     & -0.014                                        & 0.017                                          & -49.893                    & -49.922                      & -49.909                     & -0.016                             & 0.013                                          & -52.532                    & -52.547                    & -52.537                     & -0.005                           & 0.010                                          \\
$E_{dir}$         & 366.636                     & 366.640                    & 366.668                     & 0.032                                         & 0.028 & 366.478                    & 366.481                      & 366.509                     & 0.031                                         & 0.028                         & 366.063                    & 366.063                    & 366.091                     & 0.028                                         & 0.028                           \\
$E_{tot}$         & -1018.497                   & -1018.503                  & -1018.558                   & -0.061                                        & -0.055                                         & -1018.967                  & -1018.971                    & -1019.041                   & -0.074                     & -0.070                                         & -1020.546                  & -1020.546                  & -1020.733                   & -0.187                   & -0.187                                         \\
\hline
\hline
\end{tabular}
}
\end{sidewaystable}

\begin{sidewaystable}
\caption{\label{tab:Sn121er} Comparison between HFBRAD (with two mesh sizes) and HFBTHO calculations for the quasi-particle energies $E_{qp}$, neutron chemical potentials $\lambda_{n}$, neutron pairing energies $E^{n}_{pair}$, average neutron pairing gaps $\Delta_{n} $, total radii, neutron kinetic energies $E^{n}_{kin}$, total spin-orbit energies $E^{tot}_{SO}$, direct Coulomb energies $E_{dir}$, and
total energies$E_{tot}$ of the blocked one-quasiparticle states in the nucleus $^{121}$Sn. Both calculations are done with the mixed pairing. HFBTHO calculations are done by imposing spherical symmetry. The differences between HFBTHO calculations with the THO basis and HFBRAD calculations with the mesh size of 0.1 fm are also given.}
\resizebox{\linewidth}{!}{%
\begin{tabular}{l|rrrrrrrrrrrrrrr}
\hline
\hline
\multicolumn{1}{c|}{}                & \multicolumn{2}{c|}{HFBTHO $d_{3/2}$}       & \multicolumn{2}{c|}{HFBRAD $d_{3/2}$}               & \multicolumn{1}{c|}{\multirow{2}{*}{diff}} & \multicolumn{2}{c|}{HFBTHO $h_{11/2}$}      & \multicolumn{2}{c|}{HFBRAD $h_{11/2}$}              & \multicolumn{1}{c|}{\multirow{2}{*}{diff}} & \multicolumn{2}{c|}{HFBTHO $g_{7/2}$}       & \multicolumn{2}{c|}{HFBRAD $g_{7/2}$}               & \multicolumn{1}{c}{\multirow{2}{*}{diff}} \\ \cline{2-5} \cline{7-10} \cline{12-15}
\multicolumn{1}{c|}{}                & \multicolumn{1}{c|}{HO} & \multicolumn{1}{c|}{THO} & \multicolumn{1}{c|}{0.2} & \multicolumn{1}{c|}{0.1} & \multicolumn{1}{c|}{}                      & \multicolumn{1}{c|}{HO} & \multicolumn{1}{c|}{THO} & \multicolumn{1}{c|}{0.2} & \multicolumn{1}{c|}{0.1} & \multicolumn{1}{c|}{}            & \multicolumn{1}{c|}{HO} & \multicolumn{1}{c|}{THO} & \multicolumn{1}{c|}{0.2} & \multicolumn{1}{c|}{0.1} & \multicolumn{1}{c}{}              \\ \hline
\multicolumn{1}{l|}{$E_{qp}$}       & 1.283 & 1.301                   & 1.284                   & 1.284                   & -0.017                 & 1.549                  & 1.573                    & 1.55                   & 1.55                   & -0.023                                 & 2.89 & 2.901         & 2.795                   & 2.794                   & -0.106                                    \\
\multicolumn{1}{l|}{$\lambda_{n}$}   & -7.731 & -7.734                  & -7.739                  & -7.739                  & -0.005      & -7.992                  & -7.99                  & -7.986                  & -7.985                  & 0.005                            & -7.627                 & -7.628                   & -7.633                  & -7.633                  & -0.005                                    \\
\multicolumn{1}{l|}{$E^{n}_{pair}$} & -10.266 & -10.829                 & -10.511                 & -10.514                 & 0.315       & -9.522                 & -10.063                 & -9.738                  & -9.742                  & 0.321                  & -11.277                & -11.878                 & -11.491                  & -11.49                 & 0.388                                      \\
\multicolumn{1}{l|}{$\Delta_{n} $}   & 1.118 & 1.148                  & 1.132                  & 1.132                  & -0.016            & 1.092                  & 1.122                  & 1.105                  & 1.105                  & -0.016                             & 1.171 & 1.201    & 1.182                  & 1.182                  & -0.019                                     \\
\multicolumn{1}{l|}{Radius}          & 4.686 & 4.686                   & 4.686                   & 4.686                   & -0.000                  & 4.687                  & 4.687                   & 4.687                   & 4.687                   & -0.000                                    & 4.685                & 4.685                    & 4.685                   & 4.685                   & -0.000                                    \\
\multicolumn{1}{l|}{$E^{n}_{kin}$}  & 1364.466 & 1365.149                & 1364.722                & 1364.73                & -0.419                                    & 1362.315               & 1363.082                 & 1362.641                & 1362.651                & -0.431   & 1367.186                & 1367.928                & 1367.447                & 1367.451                & -0.477                                    \\
\multicolumn{1}{l|}{$E^{tot}_{SO}$} & -51.602 & -51.878                 & -51.613                 & -51.621                  & 0.257     & -50.347                & -50.684                 & -50.401                 & -50.409                & 0.275               & -54.824                & -55.107                  & -54.794                 & -54.799                 & 0.308                                    \\
\multicolumn{1}{l|}{$E_{dir}$}      & 365.921 & 365.924                 & 366.005                 & 365.952                 & 0.028         & 365.89                & 365.895                 & 365.974                 & 365.921                 & 0.026            & 366.185                & 366.188                 & 366.267                  & 366.214                 & 0.027                                     \\
\multicolumn{1}{l|}{$E_{tot}$}      & -1025.394 & -1025.482               & -1025.477               & -1025.511               & -0.029                                    & -1025.124              & -1025.192               & -1025.199               & -1025.233               & -0.041 & -1023.922              & -1024.019               & -1024.059               & -1024.093                & -0.074                                    \\ \hline
\hline
\end{tabular}
}
\end{sidewaystable}


\section{Systematic calculations on semi-magic nuclei}
\label{sec:Res}

To test the performance of the blocking approach,
firstly we have done systematic calculations on the ground states of neutron-rich semi-magic even-even nuclei and possible low-lying one-quasiparticle states in odd-$A$ semi-magic nuclei with the volume, mixed and surface pairing forces. The pairing strengths are fitted to give a 
mean neutron gap of $1.31$MeV in ${}^{120}$Sn.
The strengths are -189.2, -287.85, -517.5 MeVfm$^3$ for the volume, mixed and surface pairing forces, respectively.
 Calculations are done in a box with the size of $r=30$ fm and the step of $h=0.2$ and 0.1 fm in Sec. \ref{sec:Res} and \ref{var}, respectively. Only quasi-particle states with energy lower than 60 MeV are taken into account.
This is done in order to avoid the ultraviolet divergence problem.
No significant difference between the three pairing forces are seen for known nuclei. Only results calculated with the mixed pairing are shown below as examples. 

\begin{figure} 
\begin{center}
\includegraphics[width=0.45\textwidth]{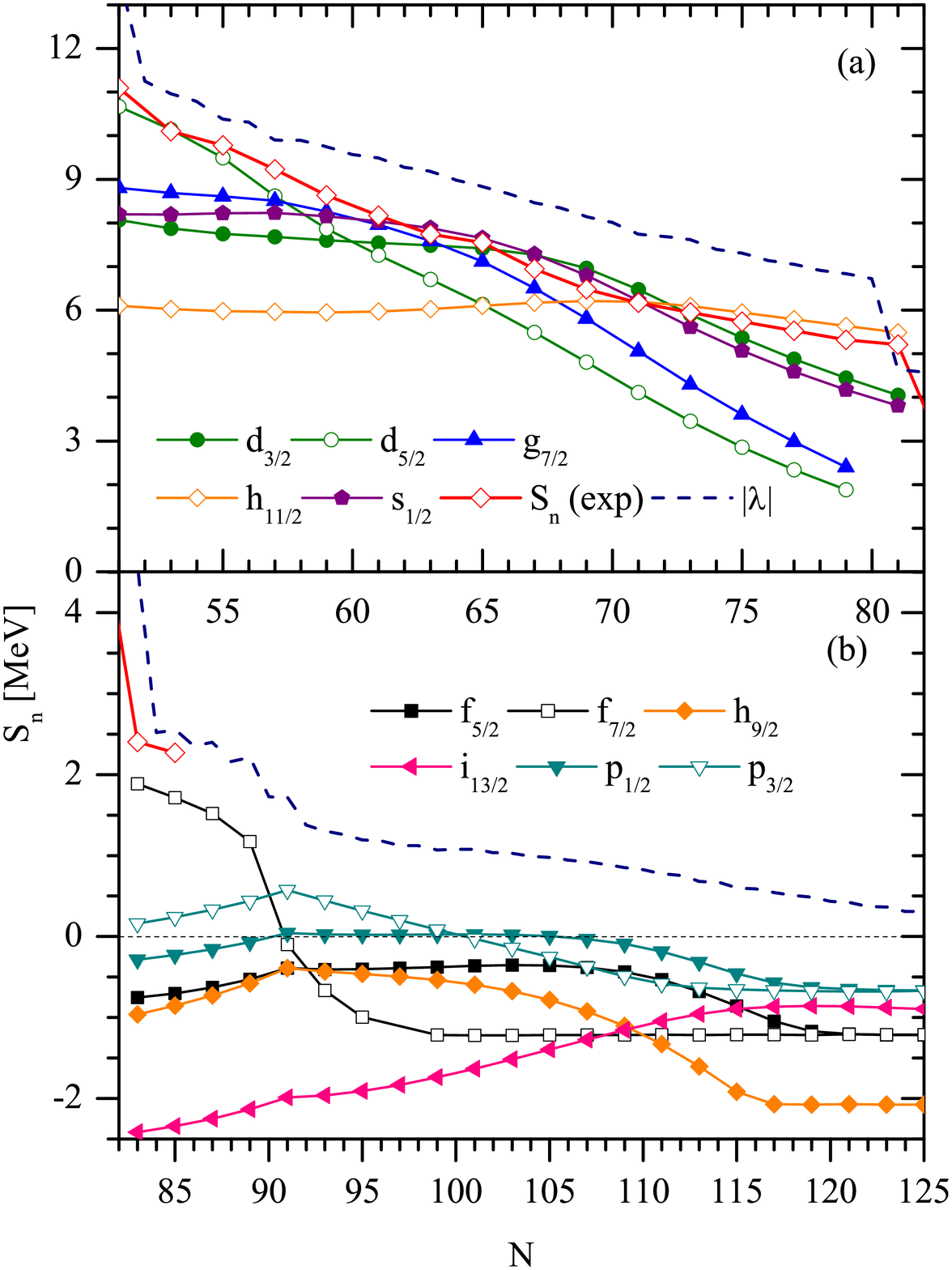}
\end{center}
\caption{\label{fig:Tin4_SN}  (color online)  The calculated one neutron separation energies of odd-$A$ Sn isotopes with different one-quasiparticle orbitals being blocked. The shown experimental data for the ground states are taken from Ref. \cite{AM12}.
}
\end{figure}

\begin{figure}  
\begin{center}
\includegraphics[width=0.45\textwidth]{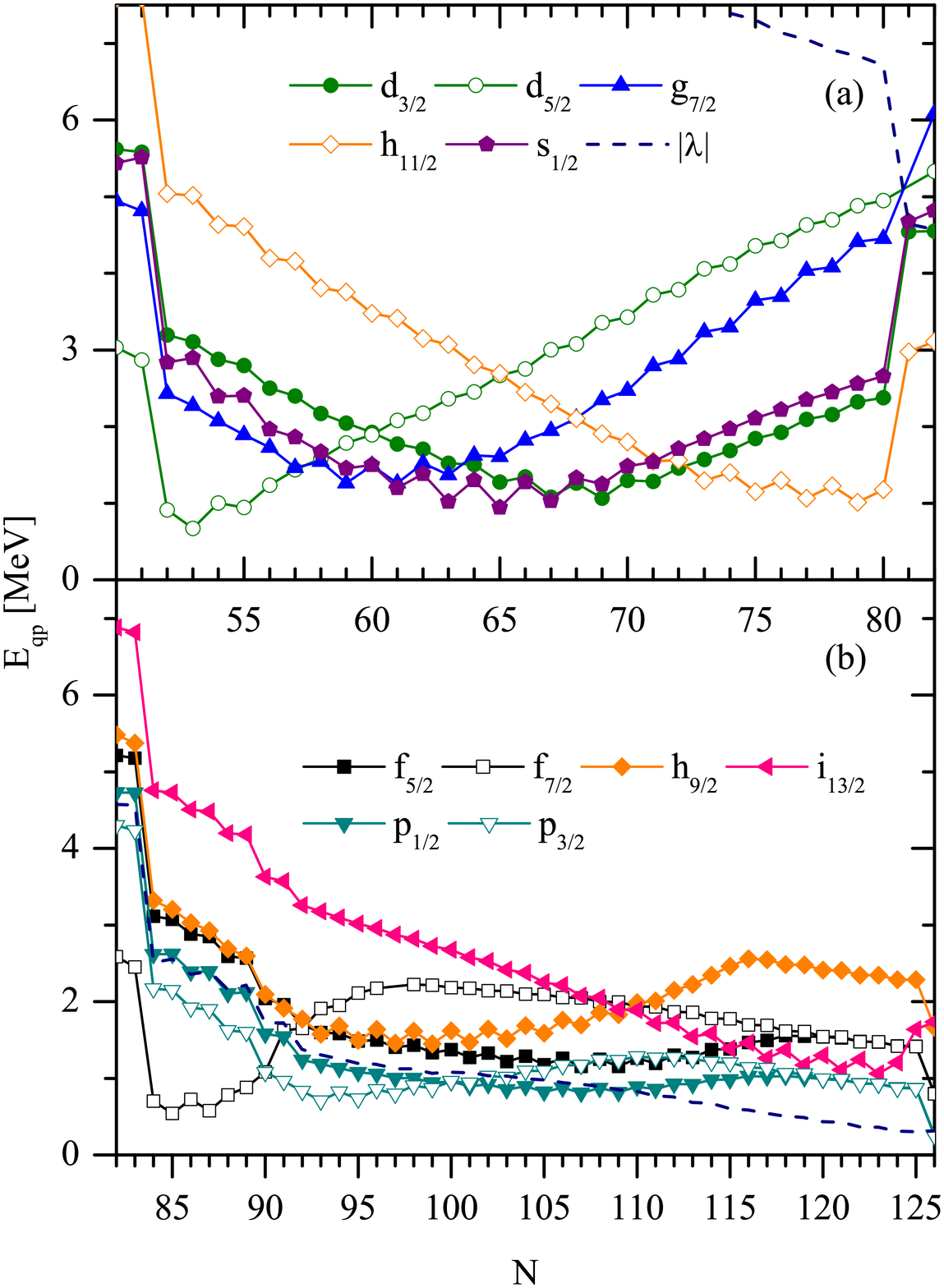}
\end{center}
\caption{\label{fig:Tin4_spe}  (color online) The calculated quasi-particle energies of different orbitals in both even and odd Sn isotopes.
}
\end{figure}

In Fig. \ref{fig:Tin4_SN} we plotted the one-neutron separation energies for the low-lying states of the odd-$A$ Sn isotopes, which is defined as
\begin{equation}
S_n(\delta)=BE(Z,N; \delta) - BE(Z,N-1),
\end{equation}
where $BE(Z,N; \delta)$ denotes the calculated (positive) binding energy of the odd-$A$ ($N$) nucleus with the quasi-particle orbital $\delta$ being blocked.
The calculated quasi-particle energies in above Sn isotopes are shown in Fig. \ref{fig:Tin4_spe} and are compared to those of their neighboring even-even isotopes. $^{131}$Sn with blocked orbital of  $1d_{5/2}$ and $0g_{7/2}$ and $^{147}$Sn with blocked orbital of $1f_{7/2}$ did not converge. 

In $^{101}$Sn the lowest single (quasi-) particle orbital is calculated to be $1d_{5/2}$ meanwhile the $0g_{7/2}$ orbital is predicted to be around 2 MeV higher. On the other hand, the recent experiment \cite{PhysRevLett.105.162502} and systematic shell model calculations \cite{PhysRevC.86.044323} tend to suggest that those two orbitals are nearly degenerate. 
There is still room for improvement for the spectroscopic quality of mean field theories, which, however, is still a challenging task \cite{PhysRevC.89.054314,PhysRevLett.113.252501,Afanasjev1409,Afanasjev4853}. 
For $^{131}$Sn, the lowest single-neutron orbital is calculated to be $0h_{11/2}$ instead of $1d_{3/2}$, even though, as shown in Ref. \cite{PhysRevLett.109.032501}, the HFB calculation can reproduce well the observed odd-even staggering in the binding energies. These two orbitals are also observed to be nearly degenerate \cite{PhysRevC.70.034312} and can be well reproduced by the nuclear shell model \cite{PhysRevC.86.044323}.

As can be inferred from both Figs. \ref{fig:Tin4_SN}(a) and \ref{fig:Tin4_spe}(a), the isotopes just above the $N=50$ are dominated by the coupling within $1d_{5/2}$ while those close to $N=82$ the lowest configuration corresponds to $0h_{11/2}$. There is a strong mixture among the $1d_{3/2}$, $0g_{7/2}$ and $2s_{1/2}$ single-particle orbitals for nuclei around $N=62$. Similarly for nuclei around $N=72$, the orbitals $0h_{11/2}$, $1d_{3/2}$ and $2s_{1/2}$ have similar quasi-particle energies. This is consistent with the fact that the one-neutron separation energies of those orbitals are also calculated to be very close to each other. One challenging task for both nuclear shell model and energy density functional is to pin down the role played by the $2s_{1/2}$ orbital in middle-shell Sn isotopes \cite{PhysRevC.86.044323} due to its low degeneracy. 
It is interesting to notice that the quasi-particle energy of the $2s_{1/2}$ orbital is calculated to be the lowest in most nuclei between $N=60$ and 67. Moreover, there is a noticeable difference between the calculated quasi-particle energies of the $2s_{1/2}$ orbitals in the odd-$A$ Sn isotopes and the neighboring even-even ones in that region. This may be related to the fact that the $2s_{1/2}$  orbital is completely blocked and does not participate in pairing correlation in the odd-$A$ isotopes.

For Sn isotopes above $N=82$, as can be seen from Figs. \ref{fig:Tin4_SN}(b) and \ref{fig:Tin4_spe}(b), the $1f_{7/2}$ and $2p_{3/2}$ are basically the only bound levels. The separation energy of the lowest $p_{1/2}$ orbital is practically zero for nuclei between $N=89$ and 109. The odd-$A$ system become unbound for $N>109$.

The neutron drip line may be defined by the chemical potential $\lambda_n$ and two-neutron separation energy, which give roughly the same results \cite{PhysRevC.91.014324,PhysRevC.91.024305}.
In Ref. \cite{PhysRevC.78.064305} it is commented that $\lambda_n$ cannot be used to define the one-neutron drip line when the HFB pairing correlation is vanishing. In our calculations as presented in Fig. \ref{fig:Tin4_SN}(b), one can see the big difference between the one-neutron separation energy and the absolute value of the chemical potential for nuclei between the positions of the predicted one-neutron and two-neutron drip lines for the corresponding odd-$A$ and even isotopes. It should also be mentioned that, as seen in Fig. \ref{fig:Tin4_spe}(b), those even-even nuclei with $N>109$ are bound but they are composed of quasi-particle continuum with $E_{qp}>\lambda$.

\begin{figure} 
\begin{center}
\includegraphics[width=0.45\textwidth]{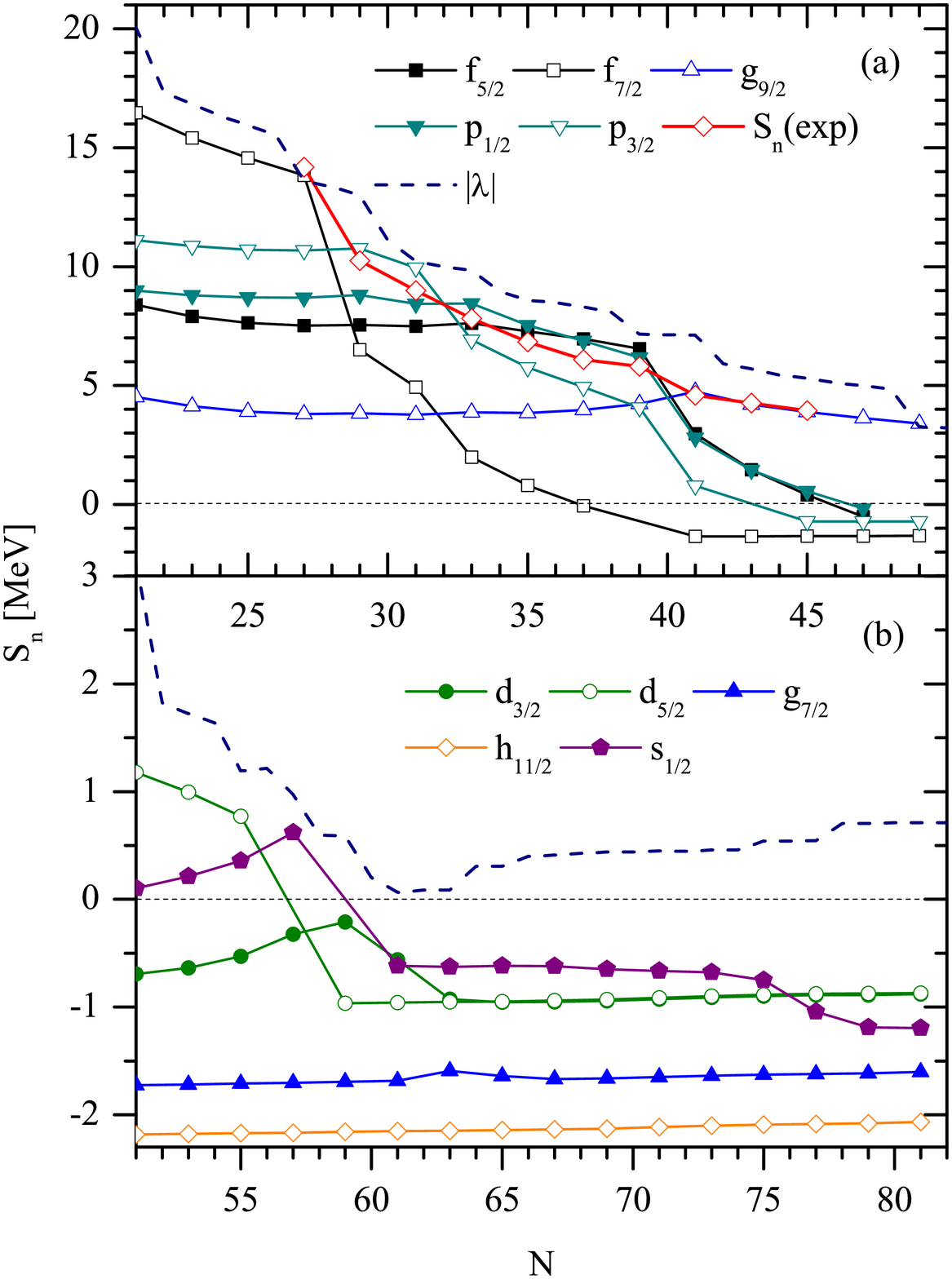}
\end{center}
\caption{\label{fig:Ni4_SN}  (color online)  Same as Fig. \ref{fig:Tin4_SN} but for Ni isotopes. The chemical potential turns positive after $N=61$.
}
\end{figure}

\begin{figure} 
\begin{center}
\includegraphics[width=0.45\textwidth]{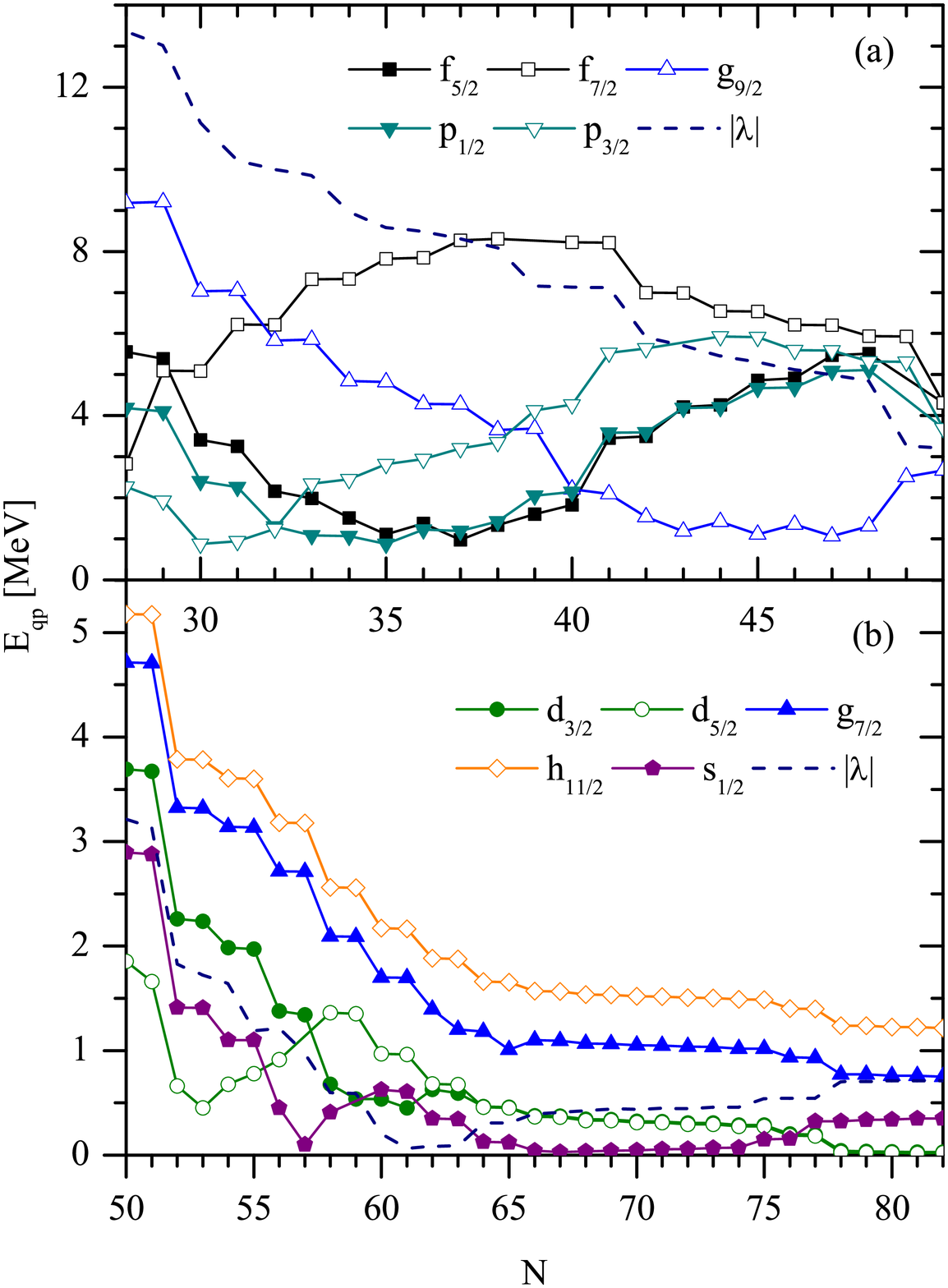}
\end{center}
\caption{\label{fig:Ni4_spe} (color online)  The calculated quasi-particle energies of different orbitals in both even and odd Ni isotopes.
}
\end{figure}

In Figs. \ref{fig:Ni4_SN} and \ref{fig:Ni4_spe} we plotted the results for Ni isotopes. In Fig. \ref{fig:Ni4_spe}(a) $^{67}$Ni with orbital $0f_{7/2}$, $^{71}$Ni with orbital $1p_{3/2}$, $^{77}$Ni with orbitals $1p_{1/2}$ and $0f_{5/2}$, and in Fig. \ref{fig:Ni4_spe}(b) $^{85}$ $^{87}$ Ni with orbitals $1d_{5/2}$ and $2s_{1/2}$ being blocked did not converge. For that isotopic chain, the two-neutron drip line is calculated to be around $N=61$. As can be seen from Fig. \ref{fig:Ni4_SN}, the $0g_{9/2}$ orbital is basically the only bound orbital as one approaches the $N=50$ shell closure from below. Beside that, one has the loosely bound $1p_{1/2}$, $0f_{5/2}$, $1d_{5/2}$ and $2s_{1/2}$ orbitals. The latter two are the only bound levels for Ni isotopes just above $N=50$ and determine the stability of those isotopes around $A=60$.

\begin{figure}
\begin{center}
\includegraphics[width=0.5\textwidth]{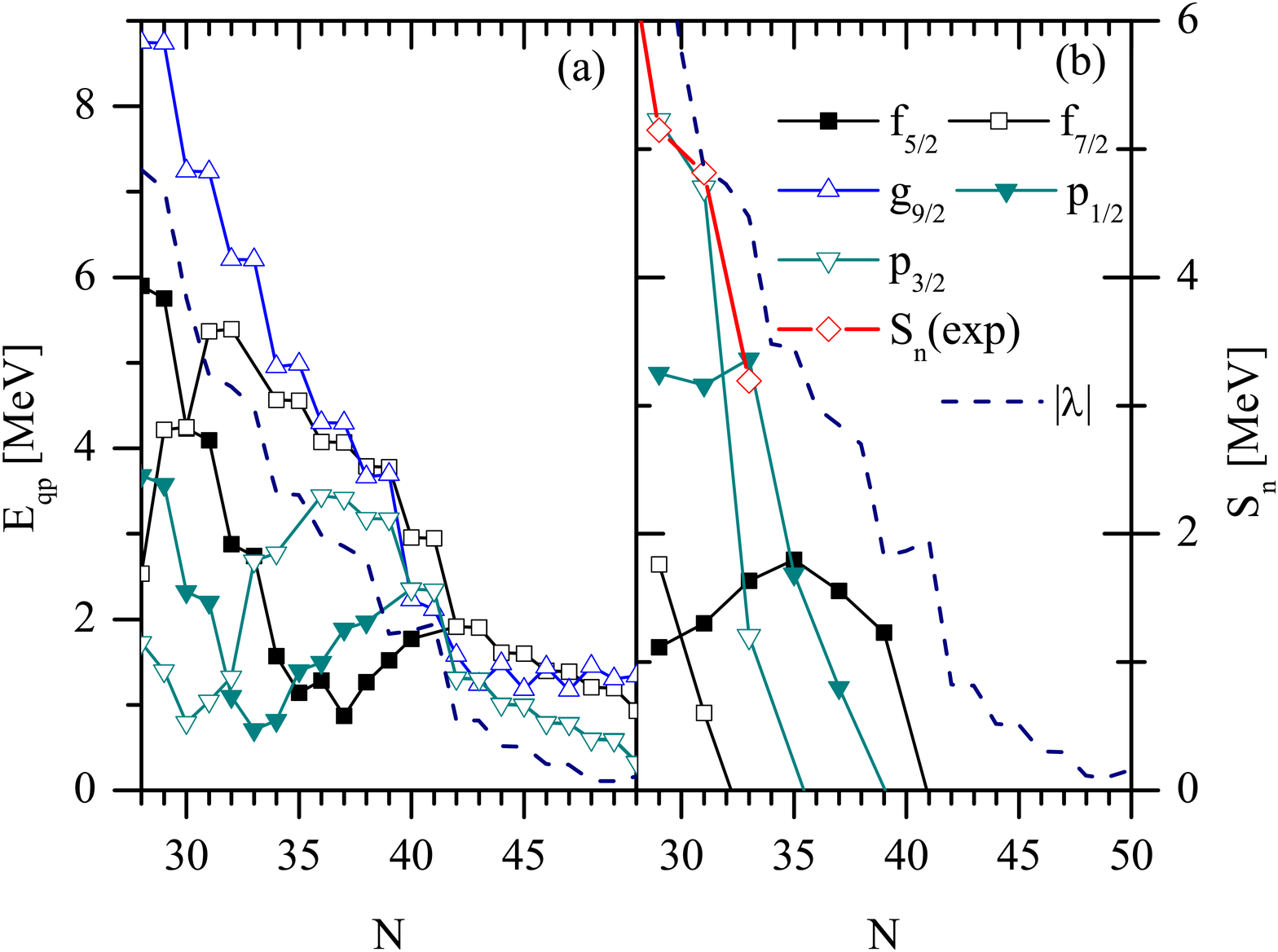}
\end{center}
\caption{\label{fig:C4_SN}  (color online) The Calculated quasi-particle energies and one neutron separation energies $S_{n}$ of different orbitals in both even and odd Ca isotopes. }
\end{figure}

The results for Ca isotopes are plotted in Fig. \ref{fig:C4_SN}. In Fig. \ref{fig:C4_SN}(a) $N=39$ and $N=41$ with blocking of $1p_{1/2}$ and $0f_{5/2}$ orbitals, respectively, did not converge. 
For odd-$A$ Ca isotopes, the one-neutron drip line is calculated to be at $N=39-41$. Meanwhile, the even-even isotopes around $N=50$ and 50 are expected to be bound as a result of the coupling to quasi-particle continuum, in particular the $l=1$ $p_{3/2}$ orbital. 
We have also done calculations for the lighter C and O isotopes. The one-neutron and two-neutron drip line for O isotopes are calculated to be at $N=17$ and $N=20$, respectively. $^{24}$O is the heaviest bound oxygen isotope that has been observed so far. The nucleus $^{25}$O has been detected to be unstable.
For the C isotopes, $^{21}$C is calculated to be the last bound odd-$A$ isotope, whereas the two-neutron drip line can be extended to $N=18$. On the experimental side, $^{21}$C is not expected to be bound \cite{Mos13,Lan85}. $^{22}$C has been expected to exhibit the neutron halo phenomenon in relation to the occupancy of the second $s_{1/2}$ orbital.  A detailed Skyrme HFB calculation on $^{22}$C was also done in Ref. \cite{Ina14} by varying the central part of Skyrme potential.

\begin{figure}
\begin{center}
\includegraphics[width=0.45\textwidth]{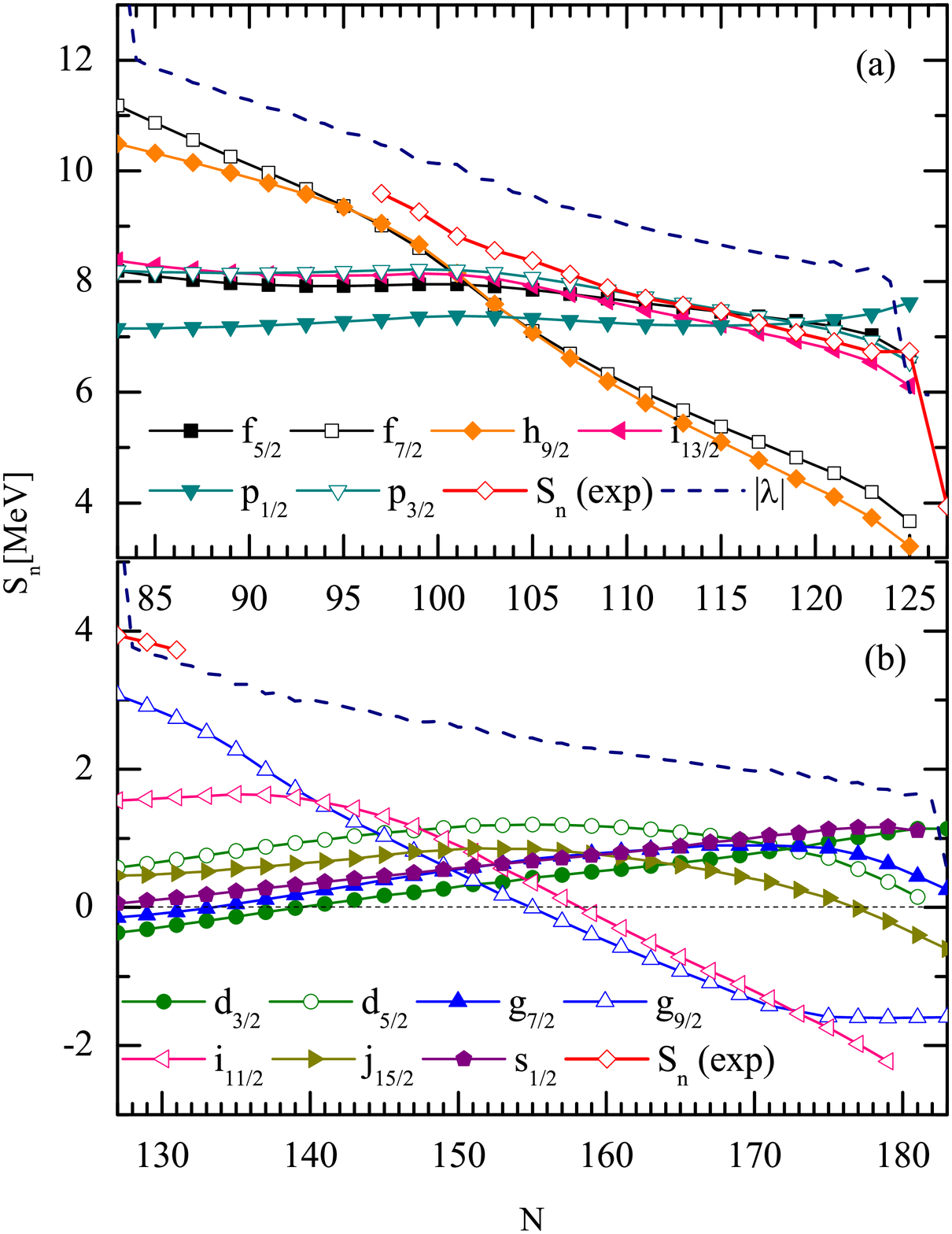}
\end{center}
\caption{\label{fig:Pb}  (color online) Same as Fig. \ref{fig:Tin4_SN} but for Pb isotopes below (a) and above (b) the $N=126$ shell closure.}
\end{figure}

The calculated one-neutron separation energies for odd-$A$ Pb isotopes are plotted in Fig. \ref{fig:Pb}. In this figure,$^{263}$Pb with orbital $0i_{11/2}$, $^{265}$Pb with orbitals $3s_{1/2}$ $2d_{5/2}$ and $0i_{11/2}$ being blocked did not converge. The ground-state one-neutron separation energies are rather well reproduced by the calculation. Noticeably differences appear around the shell closure $N=126$, where the calculation overestimated the separation energy of $p_{1/2}$ and underestimated that of $g_{9/2}$. The one-neutron and two-neutron drip lines are calculated to be near each other and are around the shell closure $N=184$.
The one-neutron drip line in Pb isotopes is much more  extended than those in lighter nuclei shown above.

Our calculations show that the low-lying one-quasiparticle states in all above semi-magic nuclei upto the neutron drip line can be evaluated by using the HFBRAD code without much numerical difficulty.
In all our results for odd-$A$ isotopes shown above and in the following, we have done blocking calculations for all possible one-neutron quasi-particle configurations. The one gives the largest binding energy (lowest HFB energy) is assigned as the ground state.

\section{Optimization of the pairing strength}
\label{var}

For calculations shown above, the strengths of the three pairing forces are determined by fitting to the pairing gap in the even-even nucleus $^{120}$Sn.
As indicated in Ref. \cite{PhysRevC.79.034306}, the global description power of the HFB calculation may be improved by fitting the strength of the pairing directly to available experimental data on OES. In Ref. \cite{PhysRevC.79.034306} two sets of HFB calculations on both spherical and deformed nuclei were done by using mixed pairing with two different pairing strengths. An optimized pairing strength was obtained by applying a linear approximation, which is slightly larger than the one determined by fitting to only $^{120}$Sn.

\begin{figure*}
\begin{center}
\includegraphics[width=0.8\textwidth]{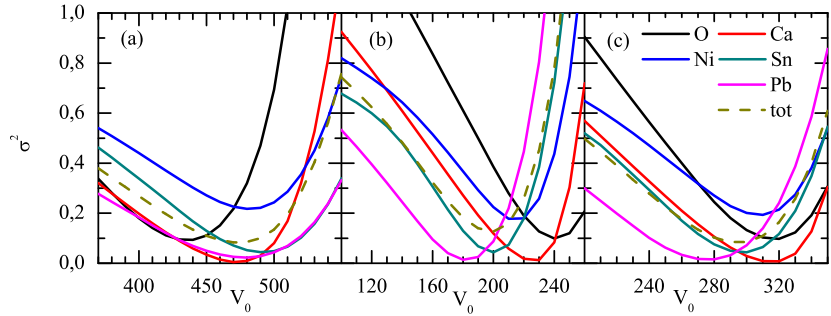}
\end{center}
\caption{\label{fig:error}  (color online) The mean square deviations (in MeV$^2$) between theory and experiment, $\sigma^2$, for the OES of semi-magic O, Ca, Ni, Sn and Pb isotopes as a function of the pairing strength (in MeVfm$^3$) for surface (left), volume (middle) and mixed (right) pairing interactions. The dashed lines are the total values while the solid lines correspond to the deviations for each isotopic chain.}
\end{figure*}
 
In Fig. \ref{fig:error} we calculated the mean square deviations between calculation and experiment, defined as $\sigma^2=\Sigma(OES_{Calc.}-OES_{Expt.})^{2}/N$ where $N$ is the number of data points considered, as a function of the strength of the pairing for available semic-magic nuclei. The OES or the empirical pairing gap is extracted from the binding energies of neighboring nuclei by using the simple three-point formula \cite{PhysRevLett.81.3599,PhysRevC.63.024308,PhysRevC.91.024305}.
As can be inferred from the figure, it may be difficult to determine the density dependence of the pairing force since the three pairing forces used predict roughly the same minimal value  for the root mean square deviation. One has $\sigma=282$ keV, 284 keV, 329 keV for the calculations with the mixed, surface and volume pairings, respectively, which correspond to pairing strengths $V_0=-295.3$, -473.4, -200.3 MeVfm$^3$. The strengths for the volume and mixed pairings thus determined are larger than those determined by fitting to the pairing gap in $^{120}$Sn whereas the strength of the surface pairing is slightly smaller than that in the latter case.

Another interesting thing is the different responses of the mean deviation to the pairing strength between light and heavy nuclei. For example, the O isotopes favors a strength that is smaller than heavy nuclei for calculations with the surface pairing. On the other hand, they favor larger pairing strengths than those of heavy nuclei for calculations with mixed and volume pairing interactions.
This is related to the fact, for calculations within the same cutoff and maximum $j$ value, calculations with the surface pairing tend to induce larger pairing gaps for light nuclei.

\subsection{Systematics of the OES}

\begin{figure}
\begin{center}
\includegraphics[width=0.45\textwidth]{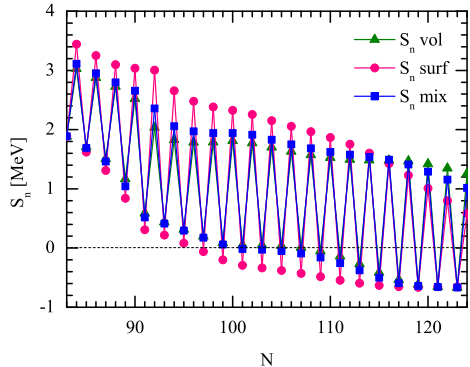}
\end{center}
\caption{\label{fig:Fit_SN} (color online) One-neutron separation energies for Sn isotopes calculated by using different pairing forces with strengths determined through global optimization.}
\end{figure}

\begin{figure}
\begin{center}
\includegraphics[width=0.45\textwidth]{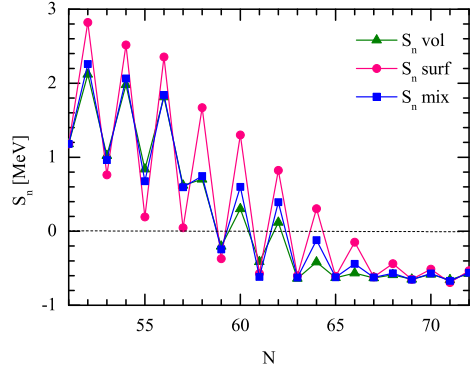}
\end{center}
\caption{\label{fig:Fit_Ni} (color online) Same as Fig. \ref{fig:Fit_SN} but for calculations on neutron-rich Ni isotopes.}
\end{figure}

We have re-calculated the one-neutron separation energies for all neutron-rich semi-magic nuclei  with the optimized pairing strengths determined above. 
In Figs. \ref{fig:Fit_SN} and \ref{fig:Fit_Ni} we plotted the results for neutron-rich Sn and Ni isotopes as examples.
In general, the choice of different density-dependent pairing forces are not expected to change noticeably the position of the two-neutron drip line in semi-magic nuclei. 
A large difference is seen between the one-neutron drip line for Sn isotopes predicted by the surface pairing calculation and those with mixed and volume pairing forces. In the latter two cases, the one-neutron drip line is at $N=99$ while those nuclei with $N$ between 101 and 109 also show a separation energy that is close to zero. For calculations with the surface pairing, the one neutron drip line clearly appears earlier at $N=95$. For Ni and Ca isotopes, the one-neutron and two-neutron drip lines may shift slightly by two neutrons between calculations with different density dependences and pairing strengths.
As for the lightest O isotopes, the one-neutron and two neutron drip lines are the same in all our calculations with the two different pairing strengths.

\begin{figure}
\includegraphics[width=0.5\textwidth]{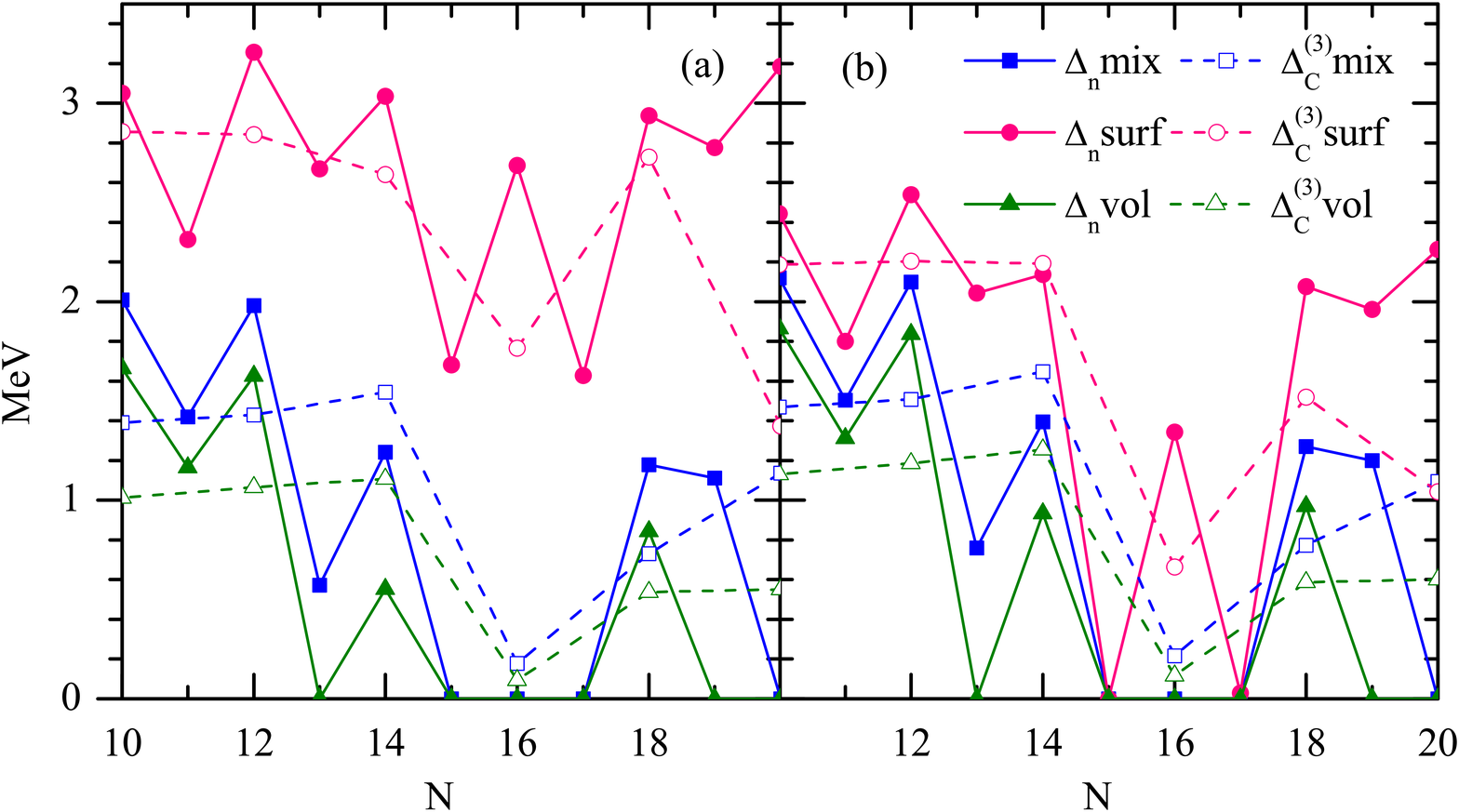}
\caption{\label{fig:Gap_O}  (color online) HFB mean pairing gaps ($\Delta_n$) and OES from the three-point formula ($\Delta_{C}^{(3)}$) in neutron-rich O isotopes calculated using mixed, volume and surface interactions with pairing strengths fitted locally to the experimental neutron gap of ${}^{120}$Sn (left) and globally to all semi-magic isotopes (right).}
\end{figure}

\begin{figure}
\includegraphics[width=0.5\textwidth]{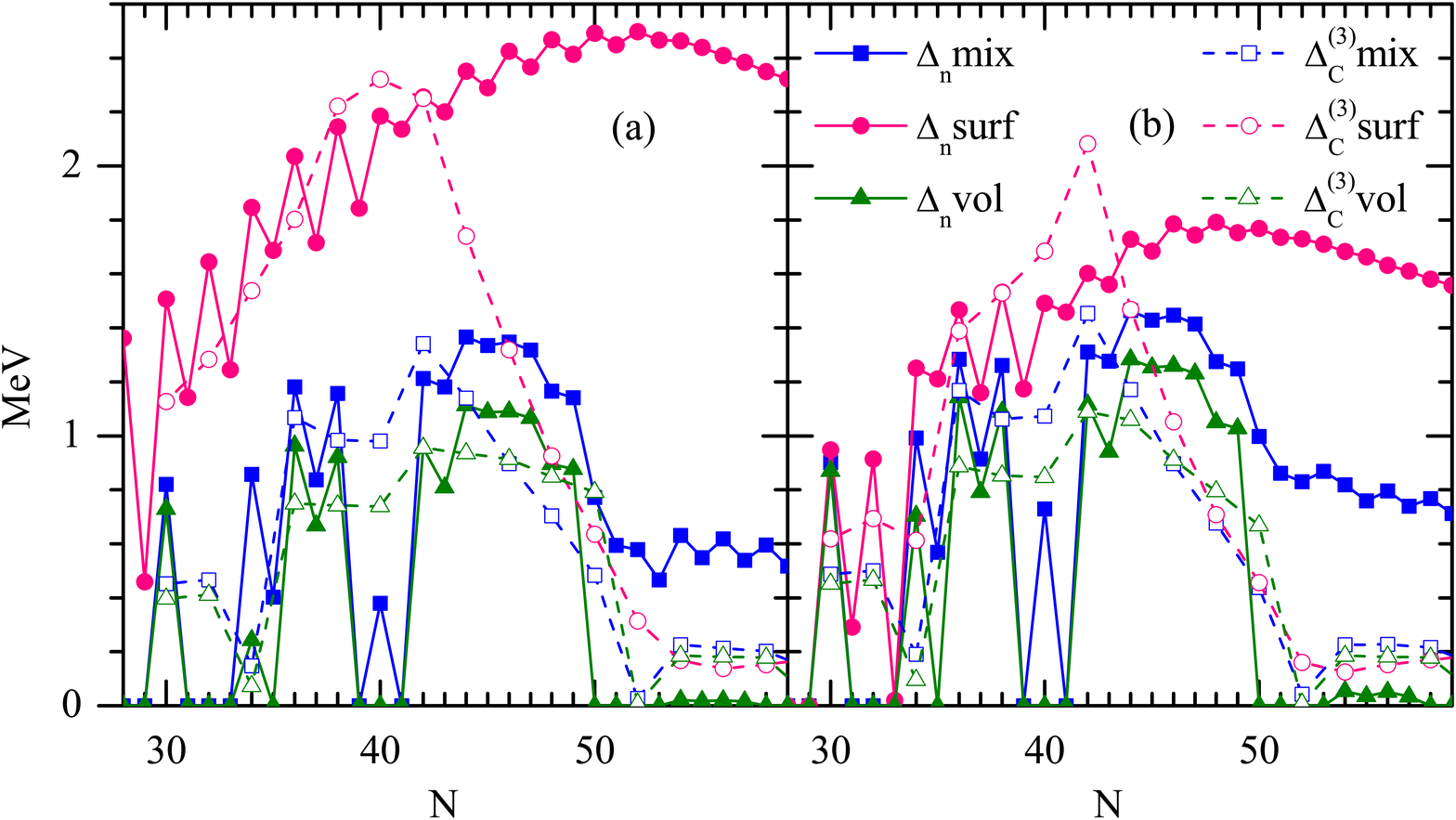}
\caption{\label{fig:Gap_Ca}  (color online) Same as Fig. \ref{fig:Gap_O} but for Ca isotopes.}
\end{figure}

\begin{figure}
\includegraphics[width=0.5\textwidth]{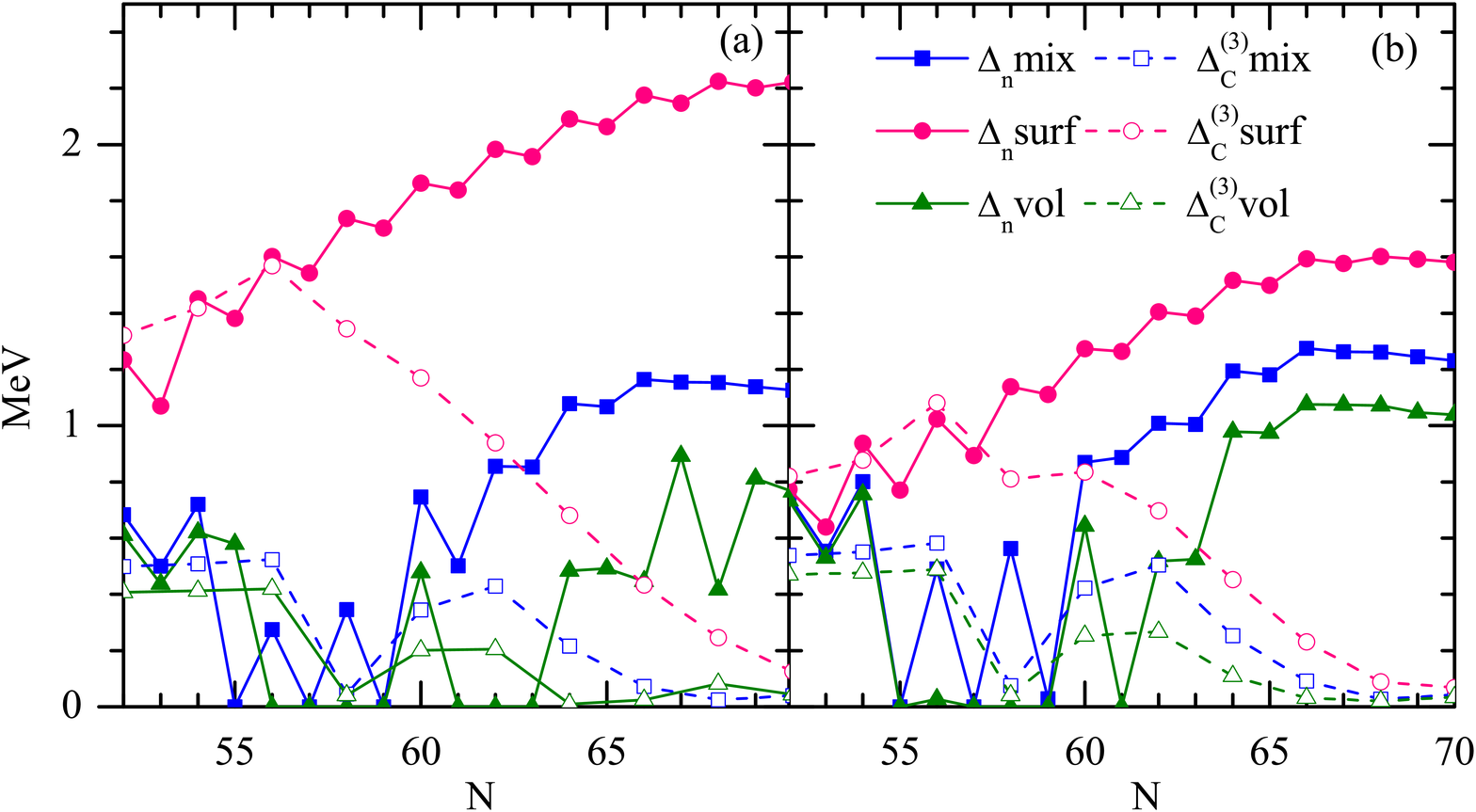}
\caption{\label{fig:Gap_Ni}  (color online) Same as Fig. \ref{fig:Gap_O} but for Ni isotopes.}
\end{figure}

\begin{figure}
\includegraphics[width=0.5\textwidth]{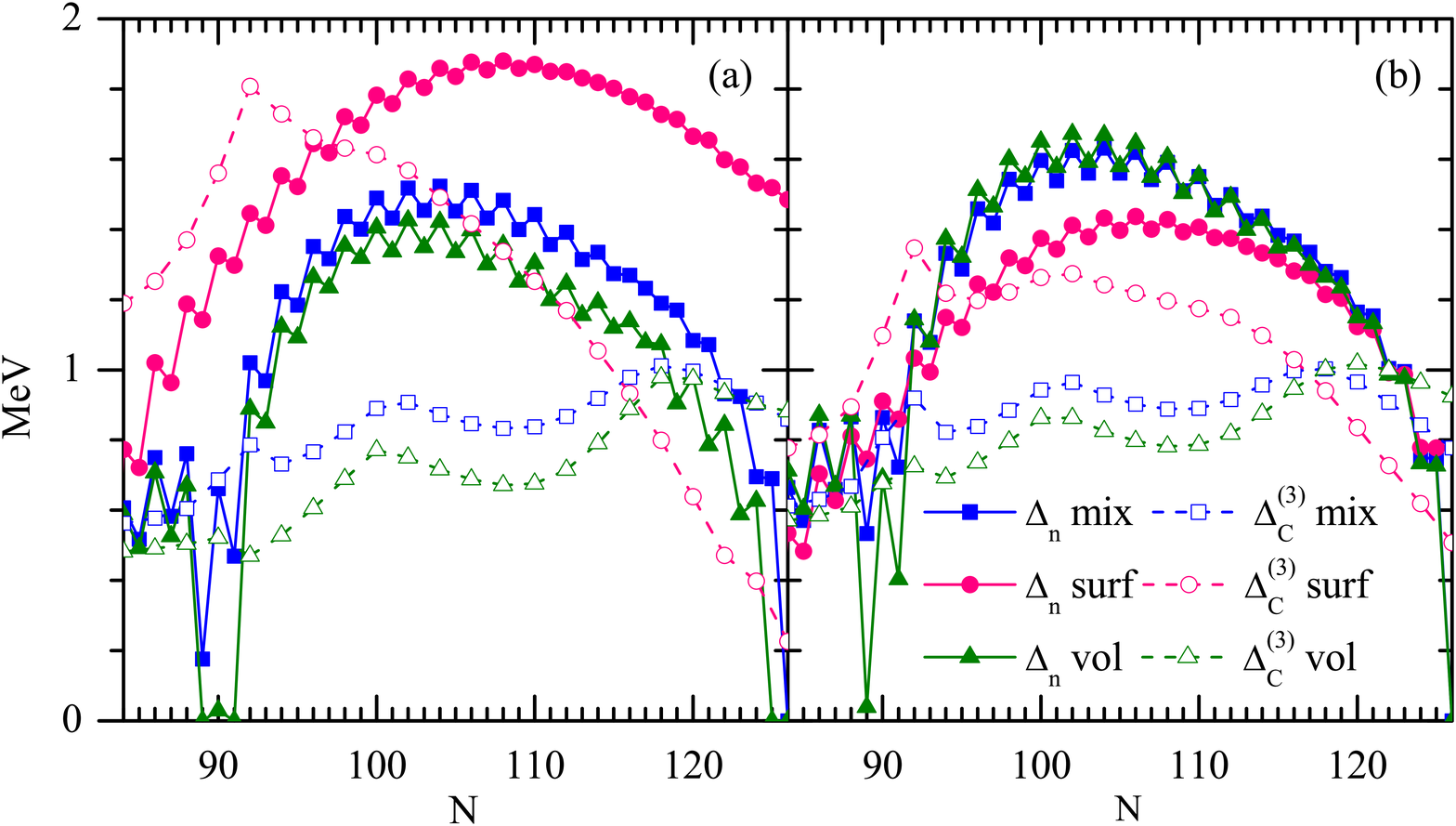}
\caption{\label{fig:Gap_Sn}  (color online) Same as Fig. \ref{fig:Gap_O} but for Sn isotopes.}
\end{figure}

\begin{figure}
\includegraphics[width=0.5\textwidth]{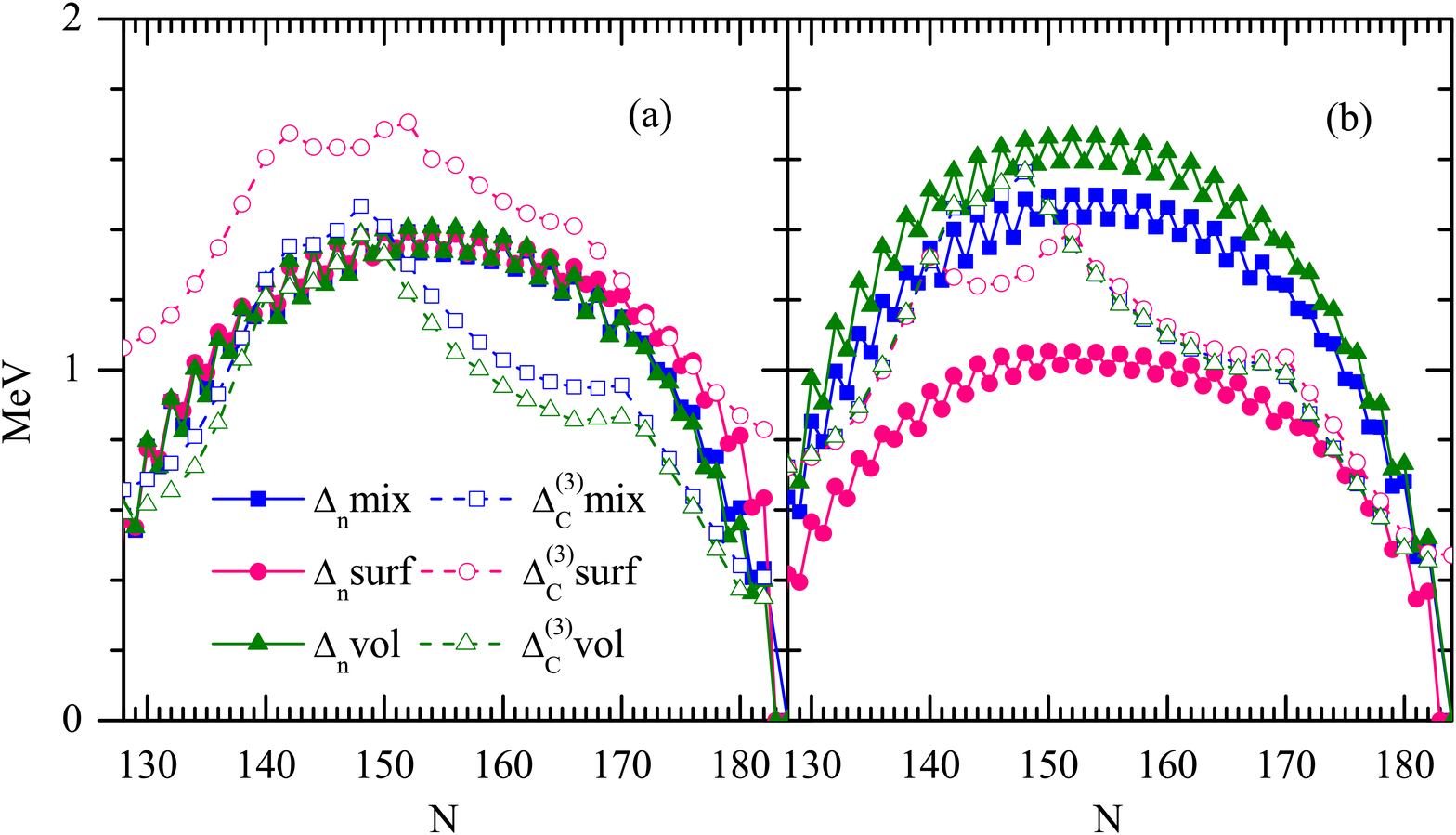}
\caption{\label{fig:Gap_Pb}  (color online) Same as Fig. \ref{fig:Gap_O} but for Pb isotopes.}
\end{figure}

As can already be inferred from Fig. \ref{fig:Fit_Ni}, the OES given by calculations with the surface pairing can be significantly larger than those from the other two pairing forces.
In Figs. \ref{fig:Gap_O} - \ref{fig:Gap_Pb} we compared the OES in above neutron-rich semi-magic nuclei  extracted from the theoretical binding energies of neighboring nuclei as calculated with the three pairing forces with the two types of pairing strength.
The results are also compared with the HFB mean gap (see, e.g., Ref. \cite{Bennaceur200596} for the definition).
Firstly it is noted that the mean gaps as predicted by the surface pairing calculation with the global optimized pairing strength are systematically smaller than those with pairing strength locally fitted to $^{120}$Sn. This is expected and is related to the fact that the pairing strength for the former calculation is slightly smaller than the latter.
Another interesting phenomenon is that,  in neutron-rich nuclei between O and Sn, noticeably large discrepancy is seen between calculations with the surface pairing and those with mixed and volume pairing forces if the 
locally fitted pairing strength is used. The difference is already seen in stable O and Ca isotopes which indicate that the surface pairing calculations for those nuclei favor a smaller pairing strength (c.f., Fig. \ref{fig:error}). 
A much smaller difference is seen in calculations with the global optimized pairing strength presented. This suggests that, in consistent with Ref. \cite{PhysRevC.80.044328}, those neutron-rich isotopes can be a sensitive probe for both the density-dependence and strength of the pairing force. For Sn isotopes, the difference already manifests itself around $N=90$ which are reachable on next generation radioactive ion beam facilities including FRIB.

The persistence of mean pairing gap around and beyond the drip line has been studied \cite{PhysRevC.88.034314}. This is more pronounced in calculations with the surface pairing than those with two other pairing forces. The OES and the mean gap are close to each other for nuclei not close to drip line. Deviations are seen for nuclei around the drip line where the OES tend to vanish. The pairing gap and the underlying correlation of the pairing wave function may be probed through pair transfer reaction instead.

A phenomenological study of the OES from measured binding energies has been done in Ref. \cite{PhysRevC.88.064329} where it was found that the OES in nuclei with $N>50$ show a decreasing behavior with increasing isospin. The OES may vanish around $N\sim 2Z$. A clear reduction of the calculated OES is seen in neutron-rich Ca, Ni and Sn isotopes with $N>2Z$. However, it may be more related to the fact that those nuclei involved are between the one-neutron and two-neutron drip ines. As for Pb isotopes, the one-neutron and two-neutron drip lines are both around $N=184$. The reduction of the mean gaps and OES there are mainly due to the shell effect.

The mean gaps and OES for Pb isotopes are relatively less sensitive to the different choice of density dependence. In particular, the OES as predicted by calculations with the globally optimized pairing strengths are practically the same for all the three pairing forces. There is a noticeable reduction of the mean gaps in calculations with the surface pairing when the globally optimized pairing strength is used.

\section{Summary}
\label{sec:con}
We have done systematic calculations on the ground states of even-even neutron-rich semi-magic isotopes as well as the low-lying one-quasiparticle states in the neighboring odd-$A$ isotopes with three different zero-range pairing forces
by solving the Hartree-Fock-Bogoliubov equations in coordinate space. The odd-$A$ nuclei are studied within the blocking scheme by assuming equal filling.
The OES in binding energy are studied in order to analyze the effect of the density dependence of the pairing force. 
The strengths of the volume, mixed and surface pairing forces are  determined in two different ways by fitting locally to the pairing gap in $^{120}$Sn and globally to available data on OES in semi-magic nuclei with $Z\geq $8. The mean deviations calculated with the three pairing forces are close to each other for the OES of known nuclei, from which one may not able to firmly pin down the density dependence of the pairing force. It is noticed that the description of the light O and Ca isotopes favors a strength that is smaller than heavy nuclei for calculations with the surface pairing. On the other hand, they favor larger pairing strengths than those of heavy nuclei for calculations with mixed and volume pairing interactions.

The choice of different density-dependent pairing forces are not expected to influence noticeably the position of the two-neutron drip line. But significant difference in the mean pairing gaps and OES can be seen between calculations with the surface pairing and the other two pairing forces. The pairing gap can persist even beyond the drip line for calculations with the surface pairing even though the OES tend to vanish. Moreover, there is a big gap between the calculated one-neutron and two-neutron drip lines in Ca and Sn isotopes. The position of the former can be sensitive to the density dependence of the pairing force.

\section*{Acknowledgement}

This work was supported by the Swedish Research Council (VR) under grant Nos. 621-2012-3805, and 621-2013-4323. The calculations were performed on resources provided by the Swedish National Infrastructure for Computing (SNIC) at NSC in Link\"oping and PDC at KTH, Stockholm.

\end{document}